\documentclass[superscriptaddress,preprint,pre,aps]{revtex4}
\usepackage[utf8x]{inputenc}
\usepackage{graphicx}
\usepackage{epstopdf}
\usepackage{amsmath}
\usepackage{amsfonts}
\begin{document}
\title{Periodic ordering of clusters and stripes in a two-dimensional lattice model. II. Results of Monte Carlo
simulation}

\author{N. G. Almarza}
\affiliation{Instituto de Qu{\'\i}mica F{\'\i}sica Rocasolano, CSIC, Serrano 119, E-28006 Madrid, Spain }
\author{J. P\c ekalski}
\affiliation{Institute of Physical Chemistry,
 Polish Academy of Sciences, 01-224 Warszawa, Poland}
\author{ A. Ciach}
\affiliation{Institute of Physical Chemistry,
 Polish Academy of Sciences, 01-224 Warszawa, Poland}
\date{\today}
\begin{abstract}
The triangular lattice model with nearest-neighbor attraction and third-neighbor repulsion, introduced 
in [J. P\c ekalski, A. Ciach and N. G. Almarza, arXiv:1401.0801 [cond-mat.soft]] is studied by Monte Carlo simulation.
Introduction of appropriate order parameters allowed us to construct a phase diagram, where different phases
with patterns made of clusters, bubbles or stripes are thermodynamically stable. We observe, in particular, 
two distinct lamellar phases - the less ordered one with global orientational order and the more ordered 
one with both orientational and translational order. Our results concern spontaneous pattern formation on 
solid surfaces, fluid interfaces or membranes that is driven by competing interactions between adsorbing 
particles or molecules.
\end{abstract}

\maketitle

\section{Introduction}

Spontaneous pattern formation in two-dimensions (2d) can occur for length scales ranging from fractions of 
a nanometer to micrometers when ions, organic molecules, macromolecules, nanoparticles
 or colloid particles are adsorbed on solid substrates, 
interfaces between two fluid phases or on biological membranes \cite{seul:93:0,seul:95:0,scheve:13:0,fan:13:0,reynaert:07:0,lenne:09:0,lang:10:0,khouri:11:0,stefaniu:12:0,thomas:12:0}.  
In the case of  membrane proteins, cluster formation enables biological functions
 that cannot be performed by single macromolecules. 
For this reason the self-assembly is very important for life processes. On the other hand, self-assembly of 
ions or nanoparticles  on surfaces or interfaces can find applications in nanotechnology or in biomimetic systems. 
 
The sum of all the effective interactions in such systems often has a form of the 
short-range attraction and long-range repulsion potential (SALR) which was intensively studied recently
~\cite{sear:99:0,pini:00:0,imperio:04:0,imperio:07:0,imperio:06:0,pini:06:0,archer:07:0,ciach:08:1,archer:08:0,bomont:12:0,
bischoff:13:0}. 
The repulsion is typically of electrostatic
origin \cite{israel:11:0,shukla:08:0,stradner:04:0,campbell:05:0},
or it is caused by polymeric brushes \cite{iglesias:12:0,panagiotopoulos:13:0}, while the attraction results from the van der Waals 
and solvent-mediated interactions
\cite{shukla:08:0,iglesias:12:0,stradner:04:0,campbell:05:0,dijkstra:99:0,scheve:13:0}. 
The size of the structural motifs  depends on the ranges of the attractive and the repulsive parts of the
 SALR potential. In the case of charged nanoparticles or  biologically relevant globular proteins in some solvents
the thickness of the clusters or stripes
is  $\sim 2-3\sigma$, where $\sigma$ is the particle diameter. 
A characteristic example is lysozyme in water \cite{shukla:08:0,kowalczyk:11:0}. 

In order to study the pattern formation in such systems, we have introduced 
 a 2d triangular lattice model with attraction between the first neighbors and repulsion between
 the third neighbors \cite{pekalski:14:0}.  In Ref.\cite{pekalski:14:0} the ground state (GS),
 the mean-field (MF) phase diagram and the 
correlation function in the Brazovskii approximation
 have been obtained.
The ground state of the model consists of the vacuum and fully occupied
 lattice for small and for large values of the chemical potential $\mu$ respectively. 
 In addition, for intermediate values of $\mu$ periodically ordered clusters, stripes or bubbles are present when
 the repulsion to attraction ratio $J^*$ is sufficiently large. 
 We have also found that at the  coexistence between the ordered phases of different 
symmetry the ground state is strongly degenerated, and the entropy per site does not vanish in the thermodynamic limit.
 The microscopic states present at the vacuum - ordered cluster phase coexistence 
consist of clusters that are located at distances larger than the repulsion range, 
but apart from this restriction the positions and the number of the clusters are
arbitrary. We interpret the collection of the disordered microstates consisting of small noninteracting 
clusters as a cluster fluid (CF)
 phase. For weak repulsion, $J^*<1$, the clusters consist of 7 particles and have a hexagonal shape, whereas for $J^*>1$
 four particles form a rhomboidal cluster.
 At the coexistence between the 
 cluster and the stripe phases 
the ground state is also strongly degenerated, and for $J^*>1$ 
 stripes of different lengths and rhomboidal clusters are packed as densely as possible
 under the constraint of no repulsion between different objects. In such microstates
one direction is distinguished, and the rotational symmetry is
 broken. We call the collection of such microstates a ``molten lamella'' (ML).

The degeneracy of the ground state must influence the phase behavior for $T>0$ (with $T$ being the temperature). However, this effect cannot be properly
described by the MF theories. The average density in the disordered phase is 
position-independent, and the contribution to the internal energy from the repulsive part of the potential 
 is an increasing function of the density. For this reason in MF
 the periodically ordered cluster phase coexists  for low $T$ with a disordered fluid of very low density, with 
 the  fraction of occupied sites $\rho\to 0$ for $T\to 0$.
In the majority of the microstates of the CF phase, however, the distances between the clusters are  
 larger than the range of the repulsion. In addition,   for $J^*>1$ the clusters are small 
and no intra-cluster repulsion is present,
therefore the repulsion contribution vanishes even for large volume fractions. Thus, beyond MF the CF phase
can be stable for quite large densities, $\rho\sim 0.2$, according to our simple estimation~\cite{pekalski:14:0}.
Similarly, the molten lamella phase 
is not present on the MF phase diagram for low $T$, although according to the GS analysis it 
 is stable at the coexistence
between the ordered cluster and stripe phases
 for $T=0$ \cite{pekalski:14:0}. 

On the other hand, for high $T$
thermal fluctuations destroy the periodic order. We have found an instability of the disordered phase with respect 
to density waves in MF. However, when fluctuations are taken into account in the Brazovskii-like 
approximation~\cite{brazovskii:75:0},
the instability with respect to the density waves is shifted to $T=0$ \cite{pekalski:14:0}. 
On the other hand, a continuous transition 
between the isotropic phase and the  phase with broken rotational symmetry 
was predicted recently for the Brazovskii functional at
 two-loop level \cite{barci:13:0}.

The above observations indicate  that
in the SALR systems the role of mesoscopic
 fluctuations is crucial for the stability of different phases consisting of aperiodic distribution of some
structural motifs. 

In this work we are interested in effects of fluctuations on pattern formation in thermal equilibrium. 
  We focus on the case of strong repulsion and perform Monte Carlo (MC)
 simulation on the triangular lattice model for $J^*=3$. Our main goal is the construction of the phase diagram. 
We should stress that  because of difficulties in the simulation procedure and in the interpretation of the results 
only  sketches of a phase diagram for the  SALR potential have been obtained 
in 2d by MC \cite{imperio:06:0} and in 3d by molecular dynamics \cite{candia:06:0}. 

In the next section the model  and the GS are described briefly. 
The MC simulations are described in sec. III.
The last section contains
 comparison between the MF and MC results, further discussion and summary.

\section{The model and its ground state}
\subsection{The model}

We assume that the particles can occupy  sites of a triangular lattice with the lattice constant equal to the 
diameter of the particles $\sigma$.
The lattice sites are ${\bf x}=x_1{\bf e}_1+x_2{\bf e}_2$, where  ${\bf e}_1$, ${\bf e}_2$ and
 $ {\bf e}_3={\bf e}_2-{\bf e}_1$ are the unit lattice 
vectors on the triangular lattice, i.e. $|{\bf e}_1|=|{\bf e}_2|=|{\bf e}_1-{\bf e}_2|=1$ (in $\sigma$-units),
 and $x_i$ are integer,  $0\le x_i\le L-1$,
where $L$ is the size of the lattice in the  directions  ${\bf e}_1$ and ${\bf e}_2$.
 We assume periodic boundary 
conditions (PBC), $L\equiv 0$ and $-1 \equiv L-1$. The number of sites on the lattices will be $M=L^2$.  

 The probability 
of a particular microscopic state
$\{\hat \rho({\bf x})\}$, where $\hat\rho({\bf x})=1(0)$ when the site ${\bf x}$ is (is not) occupied,
 has the  form 
\begin{equation}
\label{pB}
 p[\{\hat \rho({\bf x})\}]=\Xi^{-1}\exp(-\beta H[\{\hat \rho({\bf x})\}]),
\end{equation}
where $\beta=1/(k_BT)$, $k_B$ is the Boltzmann constant, 
\begin{equation}
\label{Xi}
 \Xi=\sum_{\{\hat \rho({\bf x})\}}\exp(-\beta H[\{\hat \rho({\bf x})\}])
\end{equation}
and  the Hamiltonian is given by
\begin{equation}
\label{H}
 H= \frac{1}{2}\sum_{\bf x} \sum_{\bf x'}\hat \rho({\bf x})V({\bf x}-{\bf x}')\hat \rho({\bf x'})
-\sum_{\bf x} \mu\hat \rho({\bf x}),
\end{equation}
 with the interaction energy between the cells at ${\bf x}$ and ${\bf x}+\Delta{\bf x}$

\begin{equation}
\label{V}
V(\Delta{\bf x}) = \left\{ \begin{array}{ll}
-J_1 & \textrm{for $|\Delta{\bf x}| = 1$},\\
+J_2 & \textrm{for $|\Delta{\bf x}| = 2$},\\
0 & \textrm{otherwise.}
\end{array} \right.
\end{equation}

A similar model, but  on a square lattice and with next-nearest-neighbor repulsion 
was studied in Ref.\cite{barbosa:05:0}. 
The advantage of the triangular lattice is the possibility of close packing of the particles.  
Following Ref.\cite{pekalski:13:0,pekalski:14:0}, we choose $J_1$ as the energy unit, and introduce the notation 
 $X^*=X/J_1$  for any quantity $X$ with dimension of energy.

As shown in Ref.\cite{pekalski:13:0,pekalski:14:0}, the phase diagram is symmetric with respect to the symmetry axis 
$\mu^*=\tilde V^*(0)/2=3(J^*-1)$, with $J^*=J_2/J_1$. Because of the model symmetry, we consider only 
$\mu^*\le \tilde V^*(0)/2$
 and take into account that the  sites which for $\mu^*$ are occupied (empty), 
 must be replaced by the empty (occupied)
 sites for $\mu'^* = \tilde V^*(0)-\mu^*$.

\subsection{The ground state}

\begin{figure}[ht]
\includegraphics[scale=1.08]{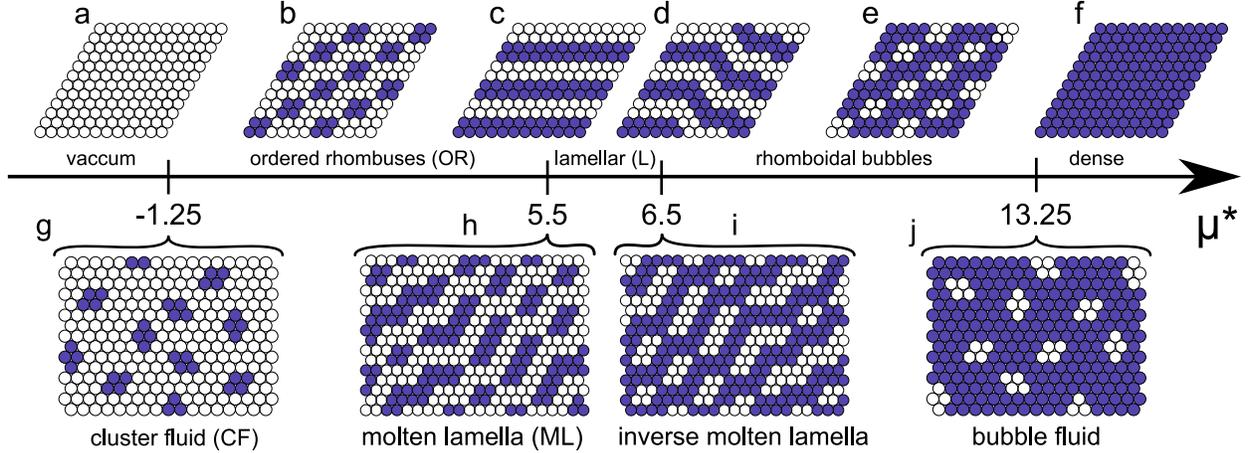}
\caption{Ground state of the model for $J^*=3$. The structures of the ordered phases are shown 
in the panels a-j, with: a) vacuum stable for $\mu^*\le -5/4$, b) ordered rhomboidal phase (OR) stable 
for $-5/4\le \mu^*\le 11/2$, 
c) and d)  lamellar phase (L) stable  for $11/2\le \mu^*\le 13/2$, e) rhomboidal bubbles 
stable  for  $ 13/2\le \mu^*\le 53/4$,
 f) dense phase stable  for  $ \mu^*\ge 53/4$. In the second row representative examples 
 of the microscopic states of the disordered 
 phases stable at the coexistences between the ordered phases are presented; 
  g)  the cluster fluid phase (CF) stable for $\mu^*=-5/4$, h) molten lamella phase (ML)
 stable for $\mu^*=11/2$, i) the inverse molten lamella stable for $\mu^*=13/2$ 
 and j) the bubble fluid phase stable for $\mu^*=53/4$.}
\label{ground_state}
\end{figure}

The $(J^*,\mu^*)$  GS has been determined in Ref.\cite{ciach:11:1,pekalski:14:0}.
 Here we limit ourselves to strong repulsion, and choose $J^*=3$.
 The  phase diagram and structure of the stable phases for  $J^*=3$ at $T^*=0$ are shown in Fig.\ref{ground_state}. 

 In order to discuss the GS, let us first consider the energy.
For strong repulsion the energy assumes a minimum when neither intra-cluster nor inter-cluster repulsion is present, 
and as many nearest-neighbors as possible are occupied. The largest cluster with no intra-cluster repulsion has the 
form of a rhombus. The energy assumes a minimum for the largest number of noninteracting rhombuses per unit area, as
 shown in Fig.\ref{ground_state}b. 

The energy competes with $-\mu N$, 
where $N$ is the number of occupied sites, and for $\mu\ll 0$ the minimum of  $H^* [\{\hat\rho({\bf x})\}]/L^2$ 
corresponds to the empty lattice (see Eq.(\ref{H})). 
The vacuum and the ordered cluster phases coexist for  $\mu^*=-5/4$. For $\mu^*=-5/4$
the disordered CF phase with a typical state shown in Fig.\ref{ground_state}g coexists with the other
two phases stable for $\mu^*=-5/4$. 
For $-5/4\le \mu^*\le 11/2$ the ordered rhombus (OR) phase is stable.
Large $\mu$ can overcompensate the effect of the repulsion, and for $\mu^*\ge 11/2$
the lamellar (stripe) phase is stable. For $\mu^*\ge 6$ the inverse phases are stable for $\mu^*$ determined 
by the symmetry of the model (see Fig.1). 
The OR and the lamellar (L) phases coexist for 
$\mu^*=11/2$. At the coexistence between the OR and the L  phases there exists
a large number of disordered states with the same value of the Hamiltonian as for the two coexisting ordered phases. 
A characteristic example of such states
 is shown in Fig.\ref{ground_state} h. 
 Such a structure  is characterized by the lack of periodicity
 in all three lattice directions, but one direction ($\hat{\bf e}_3$ in Fig.\ref{ground_state} h) is distinguished. 
We call the nonisotropic phase with noninteracting pieces of the zig-zag lamella of different length
 a 'molten lamella'.

\begin{figure}[h]
\includegraphics[scale=1]{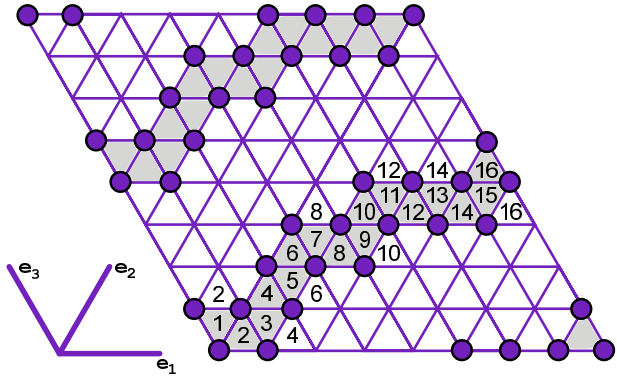}
\caption{Sketch of the construction of the ground state lamellar structures.}
\label{fig.lattice}
\end{figure}

 Note that the Hamiltonian takes the same value in the lamellar phase shown in Fig.\ref{ground_state}c,
 and in the zig-zag lamellar phase shown in Fig.\ref{ground_state}d. 
 In what follows we will describe how the GS lamellar structures can be constructed and characterized.
 This description will allow us to make an estimation of the entropy of this phase, which will be useful 
later to find the phase transitions.
Let us consider lamellas with periodicity in direction ${\mathbf e}_3$, as shown in Fig. \ref{fig.lattice}.
We start by setting three occupied positions defining an elementary 
triangle on the left end of the lattice (labelled as $1$ in Fig \ref{fig.lattice}). In order to proceed, 
 after setting this first triangle (or whatever triangle with an odd label)  we occupy a lattice position 
that closes an adjacent  triangle (chosen between those labelled with even numbers in Fig. \ref{fig.lattice},
 which are those at the top or  at the right of the previous triangle).
Then after each triangle with even index is added the next triangle has to be that on its right, since the choice of
the one on the left would break the periodicity in direction ${\mathbf e}_3$.

Once we get the reference lamellar stripe following the simple rules given above,
 it is straightforward to build up the GS lamellar structure by replicating  each piece with translations.
 The occupied sites are ${\mathbf x} ={\mathbf x}_0 + 4 k {\mathbf e}_3$,
where ${\mathbf x}_0$ is the position of the site at the reference stripe and $k>0$ is an integer number.
The structures generated in this way fulfill $\rho=1/2$, and energy $U^*/N = -2 + J^*$.

Taking into account the procedure to construct the GS lamellas it is possible
to compute the entropy of this phase at $T^*=0$. PBC add some restrictions to the building procedure. 
If we follow one lamella stripe through  PBC,  it is clear
that the number of choices of growing the reference lamella stripe in the vertical direction must be 
zero or a multiple of four. In addition, $L$ must be a multiple of four.

Taking into account the number of ways of creating lamellar structures with translational order in, at least, one
direction, and the restrictions due to PBC,  we can estimate the entropy of the system as 
\begin{equation}
S/L^2 = \frac{k_B}{L^2} \ln Q \simeq
 \frac{k_B  \ln 2}{L} + O(L^{-2}).
\label{sgs}
\end{equation}
where $Q$ is the number of ways of building up the ordered lamellar structures following the previous procedure.
 The correction term $O(L^{-2})$ includes the translational degeneracy from
the location of the triangle $1$ in Fig. \ref{fig.lattice};  the possibility of considering three directions
in the construction of the reference lamella stripe; the overcounting of straight lamellas, that
show periodicity of length four in two directions; and the restrictions due to the PBC, that
eliminate about 3/4 of the reference stripes.

\section{Monte Carlo Simulation}

Grand Canonical MC Simulations in  rhombic boxes combined
with thermodynamic integration techniques \cite{AllenTildesleyBook,FrenkelBook} have been used to sketch the 
phase diagram of the model. 

\subsection{Simulation methodology  for finite temperatures}
\label{ssec:Sim}

 Each MC move was carried out as follows: (1) A lattice position, ${\bf x}$, is chosen at random. Independently of the current state of this site,  its {\it new} state for $\hat{\rho}({\bf x})$
is chosen according to the possible interactions of a particle occupying the site ${\bf x}$. The ratio between the probabilities of choosing the two possible states is given as
\begin{equation}
\frac
{P(\hat{\rho}({\bf x})=1) }
{P(\hat{\rho}({\bf x})=0 ) } = 
\exp \left\{  \beta \left[ \mu - \sum_{\bf x'} V({\bf x}- {\bf x}') \hat{\rho}({\bf x}') 
\right]
\right\}.
\end{equation}
According to the GS analysis we expect for $J^*=3$ the presence of two ordered phases
at low temperature: lamellas with $\rho=1/2$ for values of $\mu^* \approx 6$,
and the ordered rhombus phase for $-5/4 < \mu^*  < 11/2 $. 
At high temperature, the state of the sites should be essentially
dictated by the value of the chemical potential, i.e., we will have a disordered phase
with average density given as $\bar{\rho} = e^{\beta \mu}/(1+e^{\beta \mu})$.
Then we can expect some phase transition(s) connecting the high-temperature disordered
phase with the low-temperature ordered phases.
At low temperature and  $\mu^*\approx -5/4$ we can expect to find
the disordered CF phase composed of different aggregates of particles (mainly four particle
clusters), while for increasing chemical potential the system should exhibit
a phase transition into the OR phase. Similarly, for $\mu^*\approx 11/2$ the ML phase present in the GS
is expected to be stable for $T>0$.

Taking above facts into account we have used thermodynamic integration (TI) techniques
\cite{AllenTildesleyBook,FrenkelBook,Almarza2012} to locate the phase transitions. 
These techniques are based on the numerical integration of thermodynamic potentials using simulation results.
The integration of the grand potential is carried out as
\begin{equation}
 \Omega(L, \beta,\mu_1) = \Omega(L, \beta,\mu_0) 
-  L^2 \int_{\mu_0}^{\mu_1} \check{\rho}(L, \beta,\mu) \: {\textrm d}  \mu, 
\quad \textrm{where } \check{\rho} \equiv \frac{\langle N \rangle}{L^2},
\label{ith1}
\end{equation}
when the temperature is fixed, or:
\begin{equation}
 \Omega(L, \beta_1,\mu) = \Omega(L, \beta_0,\mu) + 
 \int_{\beta_0}^{\beta_1} H (L, \beta,\mu) \: {\textrm d}  \beta,
\label{ith2}
\end{equation}
for fixed value of the chemical potential.
The values of the grand potential in the thermodynamic limit are known for
the ordered phases at low temperature, for the high-temperature limit $\beta=0$, and
for the vacuum system $\rho=0$.

In the first attempt to compute the phase diagram we run sequences of simulations
starting either from high temperatures $(\beta=0)$ or from low temperatures. 
In the latter case we used initial configurations of the corresponding stable
phases at the GS. Then by applying TI we could get a draft of the
shape of the phase diagram. We found some features in the results which
introduced some difficulty in the analysis. When considering the sequences of simulations
starting from high temperature,
we found non-monotonic behavior with the system size for different properties. 
In
addition to the expected transitions corresponding to the {\it melting}
of the GS ordered structures,  we found signatures of a number 
of additional thermodynamic transitions, which seemed to be either continuous 
or weakly first-order.  Moreover, in some cases the number of these 
apparent transitions depends on the system size: As one increases the system
size for a given value of $\mu$, the number of these transitions can increase.
The presence of these transitions, and the atypical dependence of the
results on the system size moved us to make use of parallel tempering
(or replica exchange) MC sampling \cite{Swendsen1986,Earl2005}  
in order to improve the sampling quality of the simulation.

\begin{figure}
\includegraphics[scale=1]{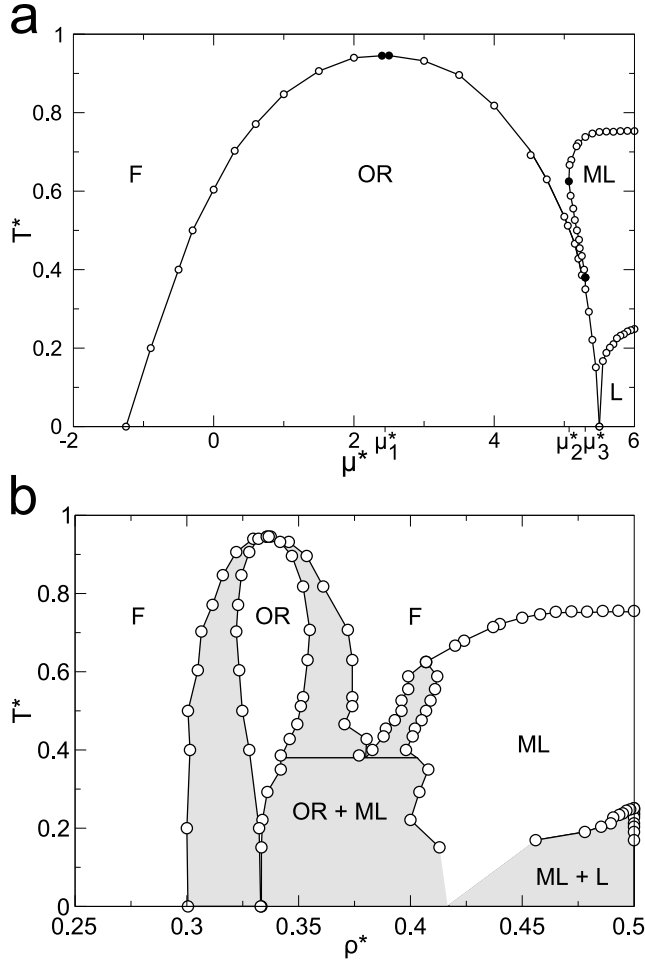}
\caption{Phase diagram in the temperature-chemical potential (a) and
in the temperature-density (b) planes. OR stands for ordered-rhombus phase,
F (fluid) for the disordered phase(s), L represents ordered lamellas, and ML
molten lamellas.
 }
\label{phd-sim}
\end{figure}

We have focused the  simulation effort 
into a relatively
large system, $L=120$, in order to get a reliable estimation of the
phase diagram. Other system sizes were considered for some representative cases, and it was found that the topology of the phase diagram did not change
and only minor displacements of the loci of the main transitions are to
be expected. The phase diagram is presented
in Fig. \ref{phd-sim}. In what follows we will address the different phase equilibria
shown therein.

\subsection{Ordered rhombus and fluid phases} 

\subsubsection{The limit of low temperature}
It is possible to compute the phase equilibria between the OR and CF phases in the
limit $T\rightarrow 0$ (for $\mu^*=-5/4)$. In this limit
 the system is 
composed exclusively by noninteracting rhomboidal clusters. In order to
analyze the possible phase transition of the system we have to pay attention to
the concavity/convexity properties of the entropy as a function of the
density of these clusters (See Refs. [\onlinecite{hoye:09:0,almarza:09:0}] 
for similar situations, in which 
a low density phase can appear  due to the presence of repulsive
interactions between particles).
In short, we can simulate an athermal lattice gas model  where the
elementary units are four-particle clusters, with exclusion rules compatible
with the lack of repulsive interactions between clusters in the lattice
 version (LSALR) of the SALR model.

This hard-core lattice gas model allowed
us to
compute the phase equilibria at conditions close to those where our standard
MC algorithm is inefficient, and it provides a consistency check
of the simulation codes by comparing its results
with those of the  
 LSALR model at low temperature.

The possible positions of a rhomboidal cluster on the triangular lattice with 
$M=L^2$ sites can be described by considering the {\it bond} connecting the two 
sites of the
rhombus that have three particles in the nearest neighbor (NN) positions. 
These {\it bond} positions  define
a Kagom\'e lattice \cite{TorquatoBook} with $3M$ sites:
Each rhombus of the LSALR model on the
triangular lattice is mapped on a site of the Kagom\'e lattice.
 Taking into account the interactions
of the LSALR model it is easy to determine the exclusion rules on the new lattice,
 i.e. which sites cannot be occupied due to the presence of a {\it particle} (occupied site).

The  simulation of the hard-core lattice model is performed using two types of moves:
translations and changes of the number of particles. Translations are carried
out by (1) Selection of a particle at random, (2) Deleting it from the system
and updating the list of non-excluded positions, and (3) Inserting it back
in one of the allowed positions (chosen at random with equal probabilities).
The second type of moves involves changes in the number of particles.
This is performed  by choosing at random with equal
probabilities either an insertion or a deletion attempt:
 If insertion is chosen and there are allowed positions, one of these
allowed positions is selected at random, and acceptance rules
are applied. If there are no allowed positions the
insertion attempt is directly rejected.
Deletion attempts are done by choosing at random a particle in the
system to be removed.

Considering detailed balance \cite{FrenkelBook} the ratio between the
acceptance probabilities, $A(N\rightarrow N')$ of an insertion and its reversal deletion move must
fulfil
\begin{equation}
\frac{A (N\rightarrow N+1) }{A(N+1\rightarrow N)} = \frac{  (N+1)  z}{N_{pos}(N)},
\end{equation}
where $z$ is the fugacity, and $N_{pos}(N)$ is the number of non-excluded
positions in the configuration with $N$ particles.
Using this simulation procedure for $L=120$, coupled with parallel tempering and 
thermodynamic integration, we obtain  the densities of the two phases
at coexistence: $\rho_F \simeq  0.30$, and $\rho_{OR}  \simeq 0.3330$. This result
is consistent with the estimate $\rho>0.2$ for the density in the CF phase 
at the coexistence with the OR phase obtained in Ref.~\cite{pekalski:14:0} beyond MF.

\begin{figure}
\includegraphics[scale=1]{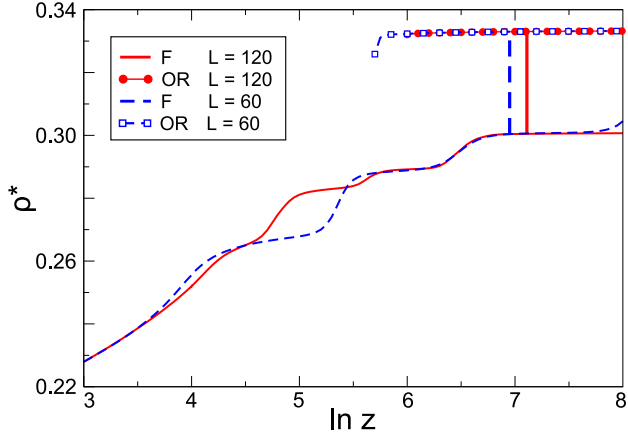}
\caption{Density as a function of the logarithm of the fugacity for the 
hard-core lattice gas model representing the F-OR equilibrium for 
$T\rightarrow 0$. Continuous
line
 represents the results for the fluid phase and $L=120$. 
Dashed lines are used for  the fluid
phase and $L=60$.
Lines with symbols mark the results for the OR phase (filled circles for $L=120$, and squares for $L=60$).
Thick vertical lines mark the estimates for the F-OR transition ($\ln z \simeq 7$ ).}
\label{hc-sim}
\end{figure}

Interesting features appear at the plots of the density of the
fluid phase as a function of $z$. For values of $z$ close to the transition,
but still in the region where the fluid phase is the stable one, the quantity:
$(\partial \rho/\partial (\ln z))$ shows an oscillatory behavior.  In the
plot of $\rho$ vs. $\ln z$  (Fig. \ref{hc-sim}) signatures of apparent weak transitions (steps-like
changes in the density) can be seen. 
The number of these steps  seems to increase
with the system size, which implies that the results for the density of the fluid phase
close to the transition depend on $L$. 
We cannot provide a definitive explanation of this behavior.
It might occur that in the fluid phase, close to the transition, the
system exhibits a short-range order, with small domains of rhombi arranged 
according to the OR structure. The size of these domains is expected to increase with
$z$, the growth of the domains might be conditioned by the effects of the PBC, which
impose some correlations which might be relevant as the correlation length of
the short-range order increases. Moreover, the use of a lattice model implies
additional spatial and orientational correlations which might amplify the effects
described above. Within this scenario we could expect that the lines $\rho$ vs $\ln z$
will become smoother for larger system sizes.

\subsubsection{F-OR transitions at finite temperature}

Sequences of simulations starting from conditions where the OR phase
is expected to be stable: $T \rightarrow 0$; $-1.25  < \mu^* < 5.50 $; showed
that this ordered phase melts irreversibly (for large system sizes, e.g. $L=120$) upon increasing $T^*$ to produce phases without long-range translational order. 
We should stress here that no evidence has been found from the simulation results
to describe the  CF and the dilute gas as thermodynamically distinct phases.
 For this reason F, instead of CF, will be used to denote disordered phase(s) at finite temperature.

In order to determine the loci of these melting transitions we firstly assumed 
that for values of $-1.25 < \mu^* \le 5.50 $ there are two possible phases,
the OR consisting of the ordered rhombuses (stable at low temperature), and a
 fluid (F) phase.
 Simulation results suggested
that the above assumption was appropriate provided that $\mu^*< \mu_2^* \simeq 5.07$.
For values in the range $\mu^* \in[\mu_2^*,5.5]$ both, the inspection
of system configurations and the use of appropriate order parameters
indicates that after the melting of the OR phase, at some temperature range, the system can adopt
 structures with some lamella character that
exhibit orientational order. We will deal with these cases later in the paper.

The loci of the OR-F transition were computed following the methodology
explained in Sec. \ref{ssec:Sim} in combination with the parallel tempering technique.
The results are shown in Fig. \ref{phd-sim}. For $\mu^* < \mu_1^* \simeq  2.4$
 the F phase is less dense than the OR phase. As one increases the chemical potential
from $\mu^*=-1.25$, the temperature at the phase equilibrium increases and for  temperatures $T^* >0.5$ the density
 of the fluid branch also increases. At $\mu^* = \mu_1^*$ the coexistence line reaches a maximum in temperature at
$T^*_2 \simeq 0.95$.
At these conditions both phases have the same density $\rho \simeq 1/3$, but
the OR-F phase transition is still first-order, with a discontinuity in the energy
\cite{almarza:09:0,hoye:09:0}.
For $\mu^* > \mu_1^*$ the density is larger 
in the F phase at F-OR coexistence.

We found that the grand canonical energy $H$, and the density $\check\rho$, for the
F phase 
as a function of the corresponding integration variable showed  
anomalies similar to those previously discussed
for the density of the model for $T\rightarrow 0$, when approaching the F-OR transition.

\begin{figure}
\includegraphics[scale=1]{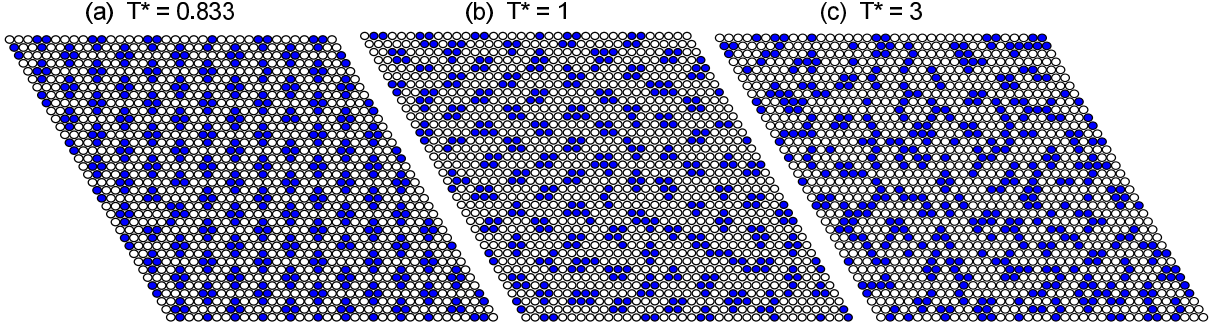}
\caption{Representative configurations for $\mu^* = 1$ with $T^* =0.833$ (OR) 
and $T^* = 1, 3$ (fluid). 
 }
\label{mrhombs}
\end{figure}

The F phase(s) show different types of structures depending on the
values of $\mu^*$, and $T$. At low temperature, $T^* \lesssim 0.5$, and
low values of $\mu^*$ the system can be described as a gas formed basically
by four-particle clusters (rhombuses, See Fig. \ref{ground_state}f), resembling the CF phase found for $\mu^*=-5/4$ at $T=0$.
 These clusters are favored by energy
considerations (sec. II B) and they will be predominant for not too low densities.
 At higher temperature and density
this phase appears as a  mixture of clusters of different sizes, which
eventually percolate as the chemical potential approaches $\mu^*=6$
\cite{TorquatoBook}. Typical snapshots of the high-$T$ OR and F phases for $\mu^* =1$ are shown in Fig.\ref{mrhombs}.

\subsection{Order parameters for lamellar phases}

In this section we introduce two order parameters with the aim
of characterizing the presence of lamellas in the medium.
In order to do this we will take into account the shape of
small fragments of the lamellas in the GS.

\subsubsection{Translational order parameter}

\begin{figure}
\includegraphics{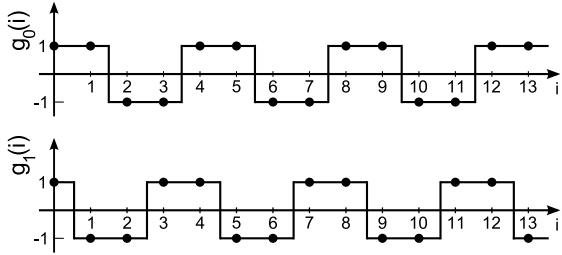}
 \caption{ The function $g_k(i)$ (see Eq.\ref{g_function})) for  $k = 0$ (top) and $k = 1$ (bottom). The lines connecting values of the function for $i = 0,1,2,\ldots$ are to guide the eyes.}
 \label{sqwe}
\end{figure}

Translational order can be checked for a given
set of simulation configurations, using the following procedure for each
of the three main directions of the lattice:
For each row $r = 0,1, \ldots, L-1$ in the selected direction $\mathbf{e}_{\alpha}$, where $\alpha = 1,2$ or $3$, 
we consider the states of its $j$ sites: 
${\hat\rho}(r,j) \equiv \hat\rho( j  \hat{\bf e}_{\alpha} +r\hat{\bf e}_{\beta})$ 
for $ j=0,1,2,\cdots,L-1$ and $\beta \neq \alpha$.
The four-site periodicity in the row $r$ in direction $\alpha$ can be checked by computing
 the parameters $p_{r \alpha}(k)$
\begin{equation}
p_{r\alpha}(k) = \frac{1}{L} \left| 
\sum_{j=0}^{L-1} \left\{ \left[ 2 {\hat \rho}\left( r,j \right) -1\right]  g_k(j) \right\} 
\right|,
\end{equation}
where $g_k(i)$ takes the form of a square wave (Fig.\ref{sqwe}) with the values
\begin{equation}
g_k(i) = \left\{ \begin{array}{lll}
1& ; & \textrm{if } \mod(i+k,4) < 2 \\
-1&; & \textrm{if } \mod(i+k,4) \ge 2,
\end{array}
\right.
\label{g_function}
\end{equation}
where $k = 0$ or $1$ and $\mod$ is the Modulo operation.
A global translational order parameter can be defined for the selected
direction, say $\mathbf{e}_{\alpha}$,  by combining the results for $k=0$ and $k=1$:
\begin{equation}
P_{\alpha} =\frac{1}{L} \sum_{r=0 }^{L-1}
\max \left[ p_{r\alpha}(0), p_{r\alpha}(1) \right].
\end{equation}
Lamellar structures with long-range translational order 
can be identified as those structures for which the order parameter P defined as
\begin{equation}
P \equiv \max [P_1,P_2,P_3],
\label{P}
\end{equation}
is close to 1.

\subsubsection{Lamellar stripes and orientational order parameter}

For chemical potentials around $\mu^*=6$, the stable configurations at low
temperature can be described as ordered lamellar phases, which exhibit long-range
translational and orientational order: Lamellar stripes {\it grow} only in two out of
three main directions of the underlying lattice (see Fig.\ref{fig.lattice}).
In order to analyze how these ordered structures are transformed,
and how the orientational order behaves on heating the system,
it is useful to define appropriate order parameters.

First, we need 
some criteria to decide if a group of particles in a given region of the system 
belongs to a lamellar stripe. 
Our criteria  works as follows:
We first look at sets of three sites forming elementary 
triangles on the lattice.
These triangles will be the units that we will characterize later as  either 
lamellar or non-lamellar. Only triangles with its three sites occupied, denoted as T$_3$, qualify to be lamellar.
Notice that these T$_3$ triangles appear also in other stable phases of the
system (See Fig. \ref{ground_state}). 
\begin{figure}[h]
\includegraphics[scale=1]{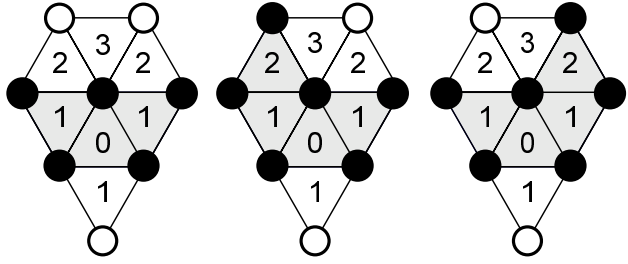}
\caption{Sketch of the definition of the lamellar triangle. Filled circles represent occupied
sites, empty circles are vacant sites.}
\label{fig.OPO}
\end{figure}
Given a  $T_3$ triangle,  we look
at the structure of its neighbor triangles. Suppose that
we pay attention to the triangle labelled as $0$ in the local
structures plotted in Fig. \ref{fig.OPO}.  The triangle $0$ can be considered
as a possible lamellar triangle if precisely two of its NN triangles
(labelled as $1$ in the figure) have their three sites occupied.
Notice that this second criterion excludes all the T$_3$ triangles in
the OR structure shown in Fig. \ref{ground_state}b. 
Finally, in order to exclude from the lamellar type those triangles
belonging to possible seven-site hexagonal clusters, we pay
attention to the triangles labelled as $2$ and $3$ in the figure. We consider
that $0$ is a lamellar triangle if,  in addition to the fulfillment of the previous conditions,
the triangle labelled as $3$ is not fully occupied.  According to these rules, 
the triangle $0$ in Fig. \ref{fig.OPO} is considered as lamellar in the three
local structures shown therein.
In addition to classifying the fully occupied triangles as lamellar or non-lamellar ones,
this method also gives the orientation of the lamellar stripe at that
point, just by considering which  two out of the three nearest-neighbor triangles are fully occupied.
There are three possible orientations which coincide with the 
principal directions on the triangular lattice.

\begin{figure}[h]
\includegraphics[scale=1]{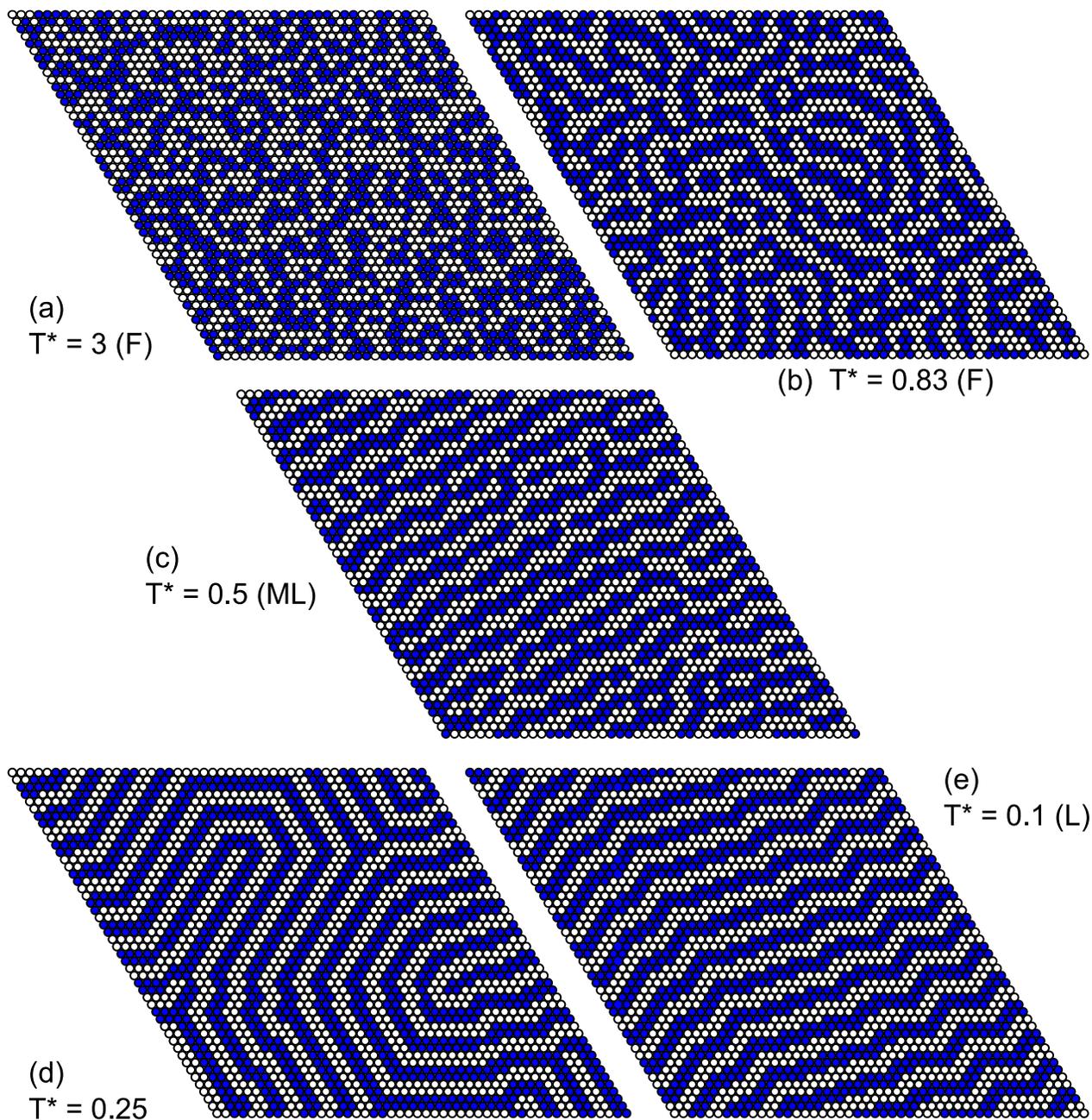}
\caption{Representative configurations for the lattice SALR model
at $\mu^*=6$ ($\langle \rho \rangle =1/2$). The system size is $L=48$. }
\label{fig.conf-lam}
\end{figure}

Visual inspection of different realizations of simulations in the neighborhood of
$\mu^* = 6$ suggests that after the melting of the ordered lamellar phase 
(the molten phase looses the translational order in the melting) 
 lamella-like stripes which show some orientational order, can  still be observed (Fig.\ref{fig.conf-lam}).
In order to quantify this behavior and relate it, eventually, with some phase
transitions we 
define appropriate order parameters.

Taking into account the ground state structures for the lamellar phases, and the 
ability of the lamellar stripes to bend (See Fig. \ref{fig.lattice} and \ref{fig.conf-lam})
 we compute separately the number of lamellar triangles that present one of the three possible orientations:
$N_{LT}^{(1)}$, $N_{LT}^{(2)}$, and $N_{LT}^{(3)}$. If the orientational order
exists, two of the orientations will be preferred by the lamellar triangles. Accordingly,
we have defined the orientational order parameter for lamellas as:
\begin{equation}
O_L = \frac{ N_{LT} -3 \left[  \min \left( N_{LT}^{(1)}, N_{LT}^{(2)}, N_{LT}^{(3)} \right)\right]  }
{N_{LT} },
\label{OL}
\end{equation}
where $N_{LT} = \sum_{i=1}^{3} N_{LT}^{(i)}$ is the total number of lamellar triangles. If no lamellar orientational order exists,
then $O_{L}$ will be close to zero, whereas it will approach $O_{L} = 1$ if the lamellar stripes
exhibit preferential orientations.

\subsection{Phase transitions of lamellas}

Parallel tempering was applied for a set of states with constant value of 
$\mu^*$, starting at $\beta=0$ and finishing 
at low temperature, typically $\beta^* = 6.0$, with the values of $\beta$ given
as $\beta^*_i= \beta_0 + i \Delta \beta$; (with $i=0,1,2,\ldots$)
and 
$\Delta \beta=0.01$
or $\Delta \beta=0.02$. Several values of $\mu^*$  were considered in the range $5.5 < \mu^* \le 6$.  
The initial configurations used for
all the replicas were those corresponding to the limit of infinite temperature and finite
$\mu^*$ (i.e. Each site of the lattice is occupied at random with probability $1/2$).
\begin{figure}[h]
\includegraphics[scale=1]{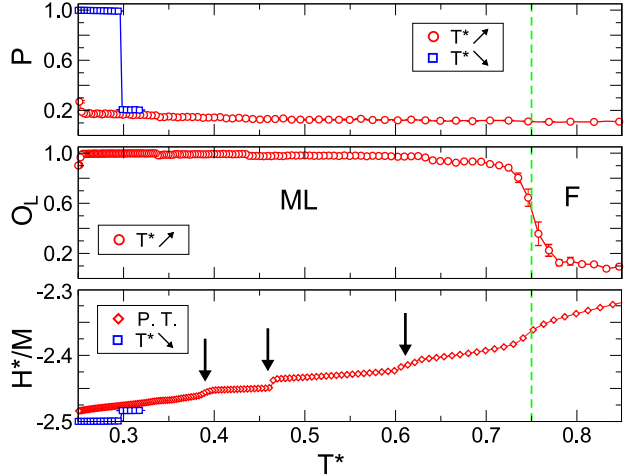}
\caption{Simulation results for the analysis of phase transitions in the lamella region.
$\mu^*=6$, $L=120$. In the top panel we plot the translational order parameter, $P$ (See Eq. (\ref{P})), 
as a function of the temperature: Circles mark the results for a sequence of simulations
starting at high temperature, squares represent the results when starting at low $T^*$ with
the initial configuration corresponding to one of the GS lamellar microstates. 
Only in the latter case the system exhibits translational order, which is lost 
when the ordered lamella melts at $T^* \simeq 0.3$. In the central panel the orientational 
order parameter for lamellas, $O_L$ (See Eq. (\ref{OL}))
is plotted for a sequence of simulations starting at high temperature: the order-disorder transition 
at $T^*\simeq 0.75$, marked with the
dashed vertical lines, can be clearly 
appreciated. Finally, in the lower panel the Grand Canonical Energy
$H^*$ per site is plotted: Diamonds mark the results obtained using the Parallel tempering scheme
starting from disordered configurations, whereas squares represent the results for the
sequence of simulations starting from the ground state configuration. The results from
Parallel tempering exhibit signatures of a likely continuous transition at the
temperature where the $O_L$ indicated an order-disorder transition. In addition some
step-like transitions (denoted by arrows) are observed below that temperature.}
\label{fig.LML6}
\end{figure}

For the ordered lamellar phase at low temperature we
 run sequences of simulations starting at low temperatures,  
typically $\beta_0^*=5.0$, using as the starting configuration one of the GS 
zig-zag lamellas, with subsequent heating steps. 
 We found that the structure of the ordered lamellar
phase hardly deviates from the GS, and therefore one can consider directly the
ideal structure of this phase and its entropy to compute $\Omega(L,\beta,\mu)$.
At some temperature the ordered lamella melts irreversibly 
(above the equilibrium melting temperature) to produce phases lacking the 
translational order,
but keeping the orientational order. In Fig. \ref{fig.LML6} we show some results for
the case of $\mu^*= 6$ and $L=120$.
By means of  TI we found that the melting
temperature for the ordered lamellas is relatively low:
$T^* \approx 0.24$ for $L=120$ and $\mu^*=6$.
This melting temperature decreases for $\mu^* < 6$ (See Fig. \ref{phd-sim}).

\begin{figure}[h]
\includegraphics[scale=1]{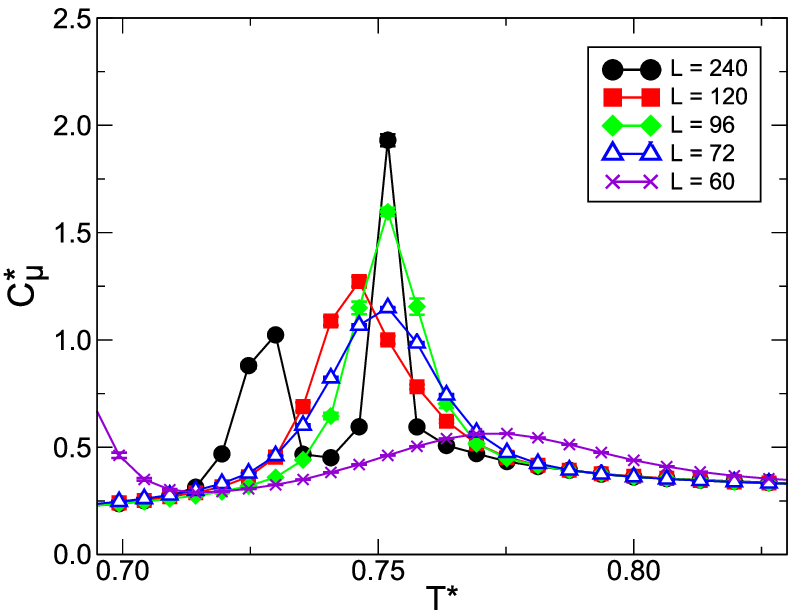}
\caption{The heat capacity at the constant chemical potential $c_{\mu}$ for $\mu^*=6.0$ and
different system sizes; $c_{\mu}$ exhibits a clear peak at the F-ML transition,
 but also non-monotonic behavior of its position and height for different system sizes.}
\label{fig.cmu}
\end{figure}

Simulation results indicate that there is a range of temperature where the configurations 
of the system can be described as molten lamellas with orientational order (See Fig. \ref{fig.LML6}). 
Looking at the variation of the order parameter $O_L$ with temperature a sharp
transition can be observed. The temperature where this transition occurs depends on $\mu$, 
and on the system size $L$. At this transition the heat capacity $c_{\mu} \equiv
L^{-2} \left( \partial H / \partial T  \right)_{L,\mu} $, presents a clear peak.
The position and the  height of this peak have non-monotonic dependences
with $L$ (in the range $24 \le L \le 120$), but the transition is observed
for all the system sizes. In spite of the non-monotonic dependence of the local maximum
value of the heat capacity $c_{\mu}^{\rm max}$(L)  with $L$, the value of $c_{\mu}^{\max}$
seems to increase slightly with $L$ (See Fig.\ref{fig.cmu}).  
These results suggest that the transition is continuous
in the range $5.5 < \mu^* \le 6.0$. A weakly first-order transition, however, cannot be ruled out.

For temperatures below this order-disorder transition, a number of apparent
 weak transitions was found. As observed for the F phase close to the
F-OR transition, the number of these transitions increases with the system
size, whereas the individual jumps in $H/M$ and $\rho$ (for $\mu^* < 6$) decrease with $L$.
Visual inspection of the configurations (Fig.\ref{fig.conf-lam}) and of the plots 
of the order parameters as a function of the temperature (Fig.\ref{fig.LML6}) did not provide  any clear conclusion about
the nature of these apparent transitions. As in the F case this behavior
could be due to effects of compatibility of the stable patterns with the PBC, and the restrictions introduced
by considering
a lattice model. Note also that the MF stability analysis predicts noninteger period $2\pi/k_b$ of the density
oscillations \cite{pekalski:14:0}, which for the lattice system together 
with PBC can lead to strong and nonmonotonic dependence on $L$.

In Fig. \ref{fig.conf-lam} we show some representative configurations for
the system at $\mu^*=6$. In panels a and b configurations of the system
in the isotropic F phase at a high temperature and temperature close to the F-ML transition respectively are shown.
 In panel b  short lamella-like stripes without any preferential orientation, 
and other cluster structures with local period of 4 in one direction, are present. 
Fig. \ref{fig.conf-lam}(c) shows a typical ML structure. Lamellar pieces 
show preferential orientations. In Fig. \ref{fig.conf-lam}(d) a metastable configuration for 
the molten-lamella phase is plotted. The structure of the lamellar stripes is quite regular but 
the translational order is not achieved and, in addition,  the global orientational order is lost.
  We have found consistently this trend for other system sizes and values of $\mu$ when the temperature 
is at (or approaching to) the region where the ML phase is metastable with respect to
the ordered L phase. Finally, in Fig. \ref{fig.conf-lam}(e) the ordered lamella structure 
is presented.

The temperature of both the F-ML and the ML-L phase transitions depends
on the chemical potential, and for both transitions assumes the maximum for $\mu^*=6$.

\subsubsection{The threshold case: $\mu^*=5.5$}
In order to understand the system behavior at the threshold value $\mu^*=5.5$
we have considered series of simulations using parallel tempering with different
system sizes. We used values of $\beta^*=1/T^* $ uniformly distributed in
the range $0 \le \beta^* <  10$. 
The simulation results showed that for system sizes $L=48, 60$,  the parallel
tempering procedure reaches, at low temperature, configurations with the
energy of GS and densities around $\rho = 5/12$, which correspond
to the ML phase shown in Figure. \ref{ground_state} h. 
For $L=120$ the simulation scheme did not reach these ground state configurations,
but the trends of the properties suggest that also in this case the  ML
 will be the stable phase (due to its large entropy) at
$\mu^*=5.5$.
The stability limits of the ML phase at low temperature
(say $T^* < 0.15$) in the phase diagrams plotted in Fig. \ref{phd-sim} 
are very hard to compute using our simulation method. Therefore
the lines plotted on the $(\rho,T)$ diagram at these conditions have
to be seen as conjectural, we cannot exclude the possibility that the ML phase might be
stable in a finite range of densities in the limit $T\rightarrow 0$.

\subsubsection{Intermediate region $\mu_2^* \le \mu^* < 5.5$} 

A more complex behavior is found when analyzing the phase behavior
as a function of the temperature at constant chemical potential in
the region $\mu_2^* \le \mu^* < 5.5$ (with $\mu_2^* \simeq 5.07$ for
$L=120$).  In spite of the fact that the
GS for this range of chemical potential is the OR phase,
on cooling the system at constant $\mu$ one finds that at some
temperature $T_{F-ML}(\mu,L)$ there is a $\textrm{F} \rightarrow \textrm{ML}$  
transition.
Depending on the value of $\mu$ this
ML phase can eventually melt again at a lower temperature, and the
isotropic phase is recovered.  Similar reentrant melting of the periodic phase was found in MF in
 the one-dimensional version of 
the present model \cite{pekalski:13:0}. This reentrant melting appears
for $\mu_2^* \le \mu \le \mu^*_3 \simeq 5.30$, and the transition
seems to be discontinuous,
with a finite change both in the Grand Canonical energy per particle: $H/M$, and in density. 
Further cooling
of the system transforms the fluid into the OR phase in a discontinuous 
transition
located through TI calculations.
In the range $\mu_3^* \lesssim \mu^* < 5.5$, no reentrant melting occurs and
a discontinuous transition between the ML and OR phases occurs at low temperature.

The reentrant behavior can be observed in the $(\mu,T)$ representation of the
 phase diagram shown in Fig.\ref{phd-sim}.
The results for $L=120$ suggest that the transition changes from a continuous 
to a weakly first order at a temperature $T^*\simeq 0.6$ (Fig.\ref{phd-sim}b), and that this
change occurs at, or close to, $\mu_2^*$, i.e. the minimum value of the chemical potential 
with the F-ML equilibrium.

\begin{figure}[h]
\includegraphics[scale=1]{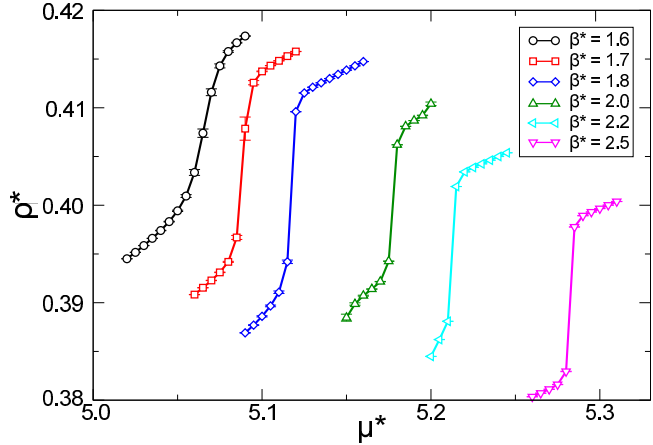}
 \caption{Density as a function of the chemical potential for different values of $\beta^*$ at
the F$\rightarrow$ML transition. }
 \label{rho_mu}
\end{figure}

The calculation of the F-ML transition below $T^* \simeq 0.6$ was carried out
by fixing the temperature and running parallel tempering MC simulation using 
several values of the
chemical potential. The plots of the density as a function of the chemical potential
indicated a continuous transition for $\beta^*=1.60$, whereas for $\beta^* \ge 1.70$,
a clear discontinuity was observed (Fig.\ref{rho_mu}). From these plots it was possible to get 
estimates for both, the chemical potential at the transition, and the density of the two phases.

The topology of the phase diagram implies the existence of a triple point 
where the OR, F, and ML phases coexist. Using the results of
simulations with $L=120$ we have estimated the locus of this point as:
$\mu^*_{TP} = \mu^*_3 \simeq 5.3$, $T^*_{TP}\simeq 0.38$.

\section{Discussion and summary}

We have used  MC simulation to compute the phase
diagram of a triangular lattice model with the SALR interactions. 
We focused on the case of strong repulsion, with the interaction potential $V({\bf x})$ 
such that $\sum_{\bf x}V({\bf x})>0$. 
The main features of the phase diagram in our model and  in the case of the SALR potential 
 with moderate repulsion in an off-lattice system \cite{imperio:06:0,imperio:07:0} are similar. 
 The details, however, are significantly different.
The most important novelty is the stability of two
different lamellar phases.

Let us  compare the MF and the MC phase diagrams for the repulsion to attraction ratio 
$J^*=3$. In the 
 $(\mu^*,T^*)$ variables the phase diagrams  (Fig. 6 in Ref.\cite{pekalski:14:0} and Fig.\ref{phd-sim}) 
are similar for low $T^*$, except that in MC the stability region of the ML phase
extends down to $T^*=0$, while in MF the $L_2$ phase appears for $T^*>0.65$. 
The relation between the ML and the L$_2$ phases obtained in MC and in MF respectively is not entirely obvious. 
In the $L_2$ phase the density oscillates in one of the lattice directions ${\bf e}_i$, 
i.e the distinguished directions in the $L_2$ and the ML phases are the same.
 The structure of the ML phase resembles the structure of the  
L phase with defects. 
The defects do not destroy the orientational order, but the translational order  is lost (see Fig.\ref{fig.conf-lam}c). 
We attribute the stability of the ML phase for $T^*\to 0$ to the effect of the degeneracy of the GS that cannot be
properly described within MF. 

For high $T^*$ the differences between the MF and MC phase diagrams in the  $(\mu^*,T^*)$ 
variables are more pronounced. 
In MF the stability region of the ordered phases extends to much higher temperatures than
found by MC simulation. Moreover, for high $T^*$ the H  (hexagonal) phase with translational 
and without orientational order of the 
rhomboidal clusters appears in the MF, but not in the MC phase diagram. 
Although we were not able to identify the H phase using MC simulation,
we cannot  rule out the physical significance of this MF result. Notice that 
in our analysis of the simulation results (see for instance Fig. \ref{hc-sim})
we found some anomalous behavior both in the equations of state 
(either $\rho(T)$ at constant $\mu$, or $\rho(\mu)$ at constant $T$), 
and in the dependence of the results on the system size in the 
putative region of the F phase close to the transition to the OR phase.
In this regard, it is interesting to point out that other models on the triangular
lattice and related models containing competing interactions 
\cite{Nakanishi1983,Dublenych:09:0,fisher1980a,selke1980}
exhibit modulated structures, and the phenomenology of Devil's staircase \cite{bak:82:0}.
Our simulation analysis cannot answer if it is the case for our lattice SALR
model, but both the behavior of the fluid phase close to the transition to the OR phase,
and the results in the region described as molten lamellas exhibit some similarities with
those  reported for this kind of systems.

Further differences between the phase diagrams concern the sequences of phase transitions for high $T^*$.
In MF the $L_2$ phase coexists with the  H phase for
high $T^*$, and extends to higher $T^*$ than the OR phase. In contrast, in MC we find that the ML phase coexists 
with the disordered F phase
for high $T^*$, and extends to lower $T^*$ than the OR phase. 
 We attribute the stability of the F phase between the OR and ML phases  to the effects of fluctuations.

In our MC simulation we do not discriminate between possibly distinct fluid F phases, despite signatures of the phase
 transitions both in the density and in the energy.
 Thus, the region marked as F in Fig.\ref{phd-sim}
 might in fact represent stability of
 a few different phases (see the snapshots in Fig.\ref{fig.conf-lam}).

The differences between the MF and MC phase diagrams are 
even stronger in the $(\rho,T^*)$ variables. The density ranges of the stability regions of the ordered phases
and the two-phase regions
are quite different in MF and in MC. 
The exception is 
the ordered L phase. It is stable for very narrow range of densities, $\rho\approx 1/2$ both in MF and in MC.

On the MC phase diagram  the ML phase coexists with the  F phase for $T^*$ higher than the temperature at the triple
point OR-F-ML, and the transition
between the two phases becomes continuous or very weakly first-order above $T^*\simeq 0.6$.
In MF  we found  a continuous phase transition between the F and the $L_2$ phases  only
for $\rho= 1/2$ \cite{pekalski:14:0}.
We have calculated the maximum $S_m$ of the structure factor $S(k)$ in the F phase
in  the Brazovskii-type approximation\cite{brazovskii:75:0}. We  
 have obtained a large value, 
$S_m\simeq 10^4$ for $T^*=0.8$.
For decreasing $T^*$ the maximum of the structure factor increases, but it diverges only for $T^*=0$. This means
a very large but finite correlation length, and an absence of the
continuous transition to the phase with periodic density for $T^*>0$ beyond MF. 
 Recently, however,  
 a continuous transition between the isotropic and the nematic phases has been predicted 
for a stripe-forming system in the Brazovskii $\varphi^4$ theory at two-loop order\cite{barci:13:0}. 
This theory is relevant for $\rho=1/2$ in our model, and supports   
the MC results for the transition between the disordered and the ML phases. 
We expect that for high $T^*$ the transition F-ML  is  continuous, but since we do not have a definite 
proof for the order of the transition,  the very weakly first-order transition is not ruled out. 

Based on the comparison of the MF and MC  phase diagrams 
we conclude that if the amplitude of the density wave in an ordered phase is small in MF, then 
 only a short-range order (such as in Fig \ref{fig.conf-lam}b)  is left in a presence of fluctuations. 
The  MF ordered phases with a large amplitude of the density oscillations  remain 
stable in a presence of fluctuations. However, instead of a coexistence of two different ordered phases,
we find a transition between each of them and a less ordered phase above a triple point.
We found the OR-ML-L triple point at $T^*=0$, and the OR-F-ML triple point at $T^*\simeq 0.38$. We may observe that 
the temperature of the triple point is  higher when the coexisting phases are less ordered.

A further interesting question is: to what extent the results for the lattice SALR
model can be extrapolated to the corresponding models in the continuum?
Regarding this point we could expect that the existence of low density ordered
phases analogous to our OR phase might strongly depend on the details of
the interaction potential. In the continuum, a low density cluster ordered phase could 
be expected  only if the
clusters formed at low temperature show little amount of polydispersity both
in a number of particles and shape.
On the other hand, the transitions involving isotropic phases and lamellar
structures seem to appear easily for systems in the continuum
\cite{archer:08:0,imperio:04:0,imperio:07:0,imperio:06:0,Reichhardt2010}. 
We think that the results presented in this
work will provide a convenient framework to analyze the possible transitions
between fluid and lamellar phases, and the possible order-disorder transition
between different lamellar phases. 

Finally, we should mention that experimental observation of thermodynamically stable patterns is 
a real challenge, and our results can guide the experimental studies. 

 \section{Acknowledgments}
The work of JP  was   realized within the International PhD Projects
Programme of the Foundation for Polish Science, cofinanced from
European Regional Development Fund within
Innovative Economy Operational Programme "Grants for innovation".
NGA gratefully acknowledges financial support from the Direcci\'on
General de Investigaci\'on Cient\'{\i}fica  y T\'ecnica under Grant No.
FIS2010-15502, from
the Direcci\'on General de Universidades e Investigaci\'on de la Comunidad de Madrid under Grant
No. S2009/ESP-1691 and Program MODELICO-CM.
We also acknowledge the financial support by the NCN grant 2012/05/B/ST3/03302

\end{document}